\documentclass[prd,preprint,showpacs,preprintnumbers,nofootinbib,eqsecnum,superscriptaddress]{revtex4}

 \usepackage[dvips,final]{graphicx}
  \usepackage{amssymb}
   \usepackage{amsmath}
    \usepackage{amsfonts}
     \usepackage{epsfig}
      \usepackage{bm}

\usepackage{mathpazo}

\usepackage[section]{placeins}

\usepackage{multirow}
\usepackage{ctable}
\usepackage{booktabs}
\usepackage{array}
\usepackage{tabularx}
\usepackage{xcolor}
\usepackage{pstricks}

\DeclareMathAlphabet{\mathpzc}{OT1}{pzc}{m}{it}
\newcommand{\sigmaefftps}{\sigma_{\rm eff,TPS}}
\newcommand{\sigmaeffdps}{\sigma_{\rm eff,DPS}}

\begin{document}

\title{Can the triple-parton scattering be observed in\\ open charm meson production at the LHC?
}

\author{Rafa{\l} Maciu{\l}a}
\email{rafal.maciula@ifj.edu.pl} \affiliation{Institute of Nuclear
Physics, Polish Academy of Sciences, Radzikowskiego 152, PL-31-342 Krak{\'o}w, Poland}

\author{Antoni Szczurek\footnote{also at University of Rzesz\'ow, PL-35-959 Rzesz\'ow, Poland}}
\email{antoni.szczurek@ifj.edu.pl} \affiliation{Institute of Nuclear
Physics, Polish Academy of Sciences, Radzikowskiego 152, PL-31-342 Krak{\'o}w, Poland}


\begin{abstract}
We investigate whether the triple-parton scattering effects can be observed in open charm production in proton-proton collisions at the LHC.
We use so-called factorized Ansatz for calculations of hard multiple-parton interactions. The numerical results for each parton interaction are obtained within the $k_{T}$-factorization approach. Predictions for one, two and three $c\bar c$ pairs production are given for $\sqrt{s}= 7$ TeV and $\sqrt{s}= 13$ TeV. Quite large cross sections, of the order of milibarns, for the triple-parton scattering mechanism are obtained. We suggest a measurement of three $D^{0}$ mesons or three $\bar{D^{0}}$ antimesons by the LHCb collaboration. Confronting our results with recent LHCb experimental data for single and double $D^{0}$ (or $\bar{D^{0}}$) meson production we present our predictions for triple meson final state: $D^{0}D^{0}D^{0}$ or $\bar{D^{0}}\bar{D^{0}}\bar{D^{0}}$. We present cross sections for the LHCb fiducial volume as well as distributions for $D^{0}$ meson transverse momentum and three-$D^{0}$ meson invariant mass. The predicted visible cross sections, including the detector acceptance, hadronization effects and $c \to D^{0}$ branching fraction, is of the order of a few nanobarns. The counting rates including $D^{0} \to K^{-}\pi^{+}$ branching fractions are given for known or expected integrated luminosities.

\end{abstract}


\maketitle

\section{Introduction}

The multi-parton scattering effects got new impulse with the start
of the LHC operation \cite{Astalos:2015ivw,Proceedings:2016tff}. There are several ongoing studies of different
processes. So far theoretical studies concentrated on double-parton
scattering. Some time ago we have shown that charm production should
be one of the best reaction to study double-parton scattering effects \cite{Luszczak:2011zp} 
(see also Ref.~\cite{Cazaroto:2013fua}).
This was confirmed by the LHCb experimental data \cite{Aaij:2012dz} and
their subsequent interpretation \cite{Maciula:2013kd,vanHameren:2014ava,Maciula:2016wci}.

Very recently also triple parton scattering was discussed in 
the context of multiple production of $c \bar c$ pairs \cite{dEnterria:2016ids}.
Inspiringly large cross sections were presented there.

We decided to verify this result within our approach which was succesfully used 
previously for single and double $D$ meson production \cite{Maciula:2013wg,Maciula:2016wci}.
Experimentally one measures rather $D$ meson (or nonphotonic leptons).
We wish to answer the question whether the triple-meson scattering
could be seen in three $D^0$ or three $\bar D^0$ production.
In order to answer the question one has to obtain cross section
for meson production, taking into account $c \to D$ hadronization
and acceptance of the existing detectors.
Reliable predictions for triple $D$ meson production should 
check consistency of model predictions for single and double
$D$ meson production with already existing experimental data. 

\section{A sketch of the model calculations}

\begin{figure}[!h]
\begin{minipage}{0.4\textwidth}
 \centerline{\includegraphics[width=1.0\textwidth]{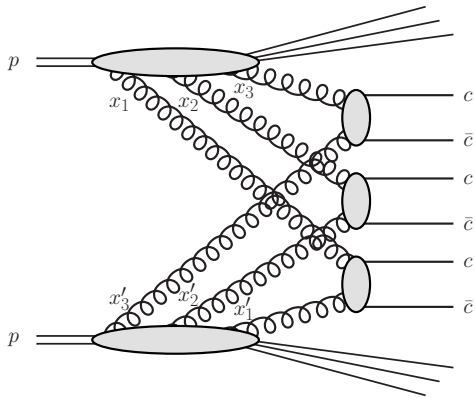}}
\end{minipage}
   \caption{
\small A diagrammatic illustration of the triple-parton scattering mechanism for triple-$c\bar c$ production in proton-proton scattering. Only the dominant at high-energies gluon-gluon fusion partonic subprocesses are taken into account.
 }
 \label{fig:diagram}
\end{figure}

The triple-parton scattering (TPS) mechanism for $pp \to c\bar c c\bar c c\bar c X$ reaction is schematically illustrated in Fig.~\ref{fig:diagram}.
The corresponding inclusive TPS cross section in a general form \cite{Calucci:2009ea,Maina:2009sj,Snigirev:2016uaq} can be written as follows:
\begin{eqnarray} 
\label{hardAB}
\sigma^{\rm TPS}_{pp \to c \bar c c \bar c c \bar c }  &=& \left(\frac{1}{3!}\right) \int \; \Gamma^{ggg}_{p}(x_1, x_2, x_3; {\vec b_1},{\vec b_2}, {\vec b_3}; \mu^2_1, \mu^2_2, \mu^2_3)\nonumber \\
& & \quad \times \quad \hat{\sigma}_{c\bar c}^{gg}(x_1, x'_1,\mu^2_1) \; \hat{\sigma}_{c\bar c}^{gg}(x_2, x'_2,\mu^2_2) \; \hat{\sigma}_{c\bar c}^{gg}(x_3, x'_3,\mu^2_3)\nonumber\\
& & \quad \times \quad \Gamma^{ggg}_{p}(x'_1, x'_2, x'_3; {\vec b_1} - {\vec b},{\vec b_2} - {\vec b},{\vec b_3} - {\vec b}; \mu^2_1, \mu^2_2, \mu^2_3)\nonumber\\
& & \quad \times \quad dx_1 \; dx_2 \; dx_3 \; dx'_1 \; dx'_2 \; dx'_3 \; d^2b_1 \; d^2b_2 \; d^2b_3 \; d^2b,
\end{eqnarray}
where $x_{i}$, $x'_{i}$ are the longitudinal momentum fractions, $\mu_{i}$ are the renormalization/factorization scales,
 $\hat{\sigma}_{c\bar c}^{gg}(x_i, x'_i,\mu^2_i)$ are the partonic cross sections for $gg\to c\bar c$ mechanism and $\frac{1}{3!}$ is the combinatorial factor relevant for the case of the three identical final states. 
The above TPS hadronic cross section is expressed in terms of the so-called triple-gluon distribution functions
$\Gamma^{ggg}_{p}(x_1, x_2, x_3; {\vec b_1},{\vec b_2}, {\vec b_3}; \mu^2_1, \mu^2_2, \mu^2_3)$ that contain additional
informations about positions ${\vec b_{i}}$ of the three corresponding partons in the transverse plain of the colliding protons.

The triple parton distribution functions (triple PDFs) shall account for all possible correlations between the partons, not only kinematical and spatial one, but also including spin and color/flavor correlations. The MPI theory in this general form is well established (see \textit{e.g.} Refs.~\cite{Diehl:2011tt,Diehl:2011yj}) but not yet fully applicable for phenomenological studies. The objects like triple PDFs (and even double PDFs in the case of DPS) are under intense theoretical studies but their adoption
to real process calculations is still limited.

Therefore, in practice one usually follows the so-called factorized Ansatz, where the correlations between partons are neglected and longitudinal and transverse degrees of freedom are separated. According to these approximations the triple-gluon PDFs from Eq.~(\ref{hardAB}) take the following form:  
\begin{eqnarray} 
\label{DxF}
\Gamma^{ggg}_{p}(x_1, x_2, x_3; {\vec b_1},{\vec b_2}, {\vec b_3}; \mu^2_1, \mu^2_2, \mu^2_3)
= D^{ggg}_p(x_1, x_2, x_3; \mu^2_1, \mu^2_2, \mu^2_3) F({\vec b_1}) F({\vec b_2}) F({\vec b_3}),
\end{eqnarray} 
where $D^{ggg}_p(x_1, x_2, x_3; \mu^2_1, \mu^2_2, \mu^2_3) = g(x_1; \mu^2_1) g(x_2; \mu^2_2) g(x_3; \mu^2_3)$ is the product of single gluon PDFs
and $F({\vec b_i})$ describe the gluon distributions in transverse plane. The transverse factors $F({\vec b_i})$ of the two triple-gluon PDFs in Eq.(\ref{DxF}) are connected to the proton-proton overlap function
via the following relation:
\begin{equation} 
\label{f}
\mkern-14mu T({\vec b}) = \int F({\vec b_i}) F({\vec b_i -b})\;d^2b_i ,\,\; {\rm with}\,\int d^2b \; T({\vec b})=1,
\end{equation} 
and are usually assumed to be universal for all types of partons.

Taking all together, the formula for inclusive TPS cross section (Eq.~(\ref{hardAB})) can be simplified to the pocket form:
\begin{equation} 
\label{doubleAB}
\sigma_{pp \to c \bar c c \bar c c \bar c }^{\rm TPS} =  \left(\frac{1}{3!}\right)\, \frac{\sigma_{pp \to c \bar c}^{\rm SPS} \cdot
\sigma_{pp \to c \bar c}^{\rm SPS} \cdot \sigma_{pp \to c \bar c}^{\rm SPS}}{\sigmaefftps^2},
\end{equation} 
where the triple-parton scattering normalization factor $\sigmaefftps$ is related to the overlap function from Eq.~(\ref{f}) via the following expression:
\begin{eqnarray} 
\sigmaefftps^2=\left[ \int d^2b \,T^3({\vec b})\right]^{-1}\,.
\label{eq:sigmaeffTPS}
\end{eqnarray} 
The normalization factor $\sigmaefftps$ contains all unknowns about the TPS dynamics. Its pure geometrical interpretation comes
from the practical approximations of the factorized Ansatz mentioned above. In principle, taking into account various parton correlations as well as multi-parton PDF sum rules \cite{Golec-Biernat:2015aza} or including perturbative-parton-splitting contributions \cite{Ryskin:2011kk,Gaunt:2012dd, Gaunt:2014rua} may lead to a breaking of the pocket-formula. However, most of the violation sources are expected to vanish for processes driven by small-$x$ partons (see \textit{e.g.} Refs.~\cite{Kasemets:2014yna,Echevarria:2015ufa}), that is exactly the case of charm production at high energies. As was shown by us, \textit{e.g.} in Ref.~\cite{vanHameren:2014ava}, the factorized framework seems to be sufficient to explain the LHCb double charm data. Therefore, we think that it can be safely used, at least as a starting point, to draw practical conclusions also in the case of triple charm production.

In principle, the DPS normalization factor $\sigmaeffdps$ was extracted experimentally from several Tevatron and LHC measurements (see \textit{e.g.} Refs.~\cite{Astalos:2015ivw,Proceedings:2016tff} and references therein) and its world average value is $\sigmaeffdps \simeq 15 \pm 5$ mb\footnote{A detailed study of the $\sigmaeffdps$ can be found in Ref.~\cite{Seymour:2013qka}.}. Such experimental inputs are not available for $\sigmaefftps$ in studies of triple-parton scattering. However, as was shown in Ref.~\cite{dEnterria:2016ids} for proton-proton collisions, the latter quantity can be expressed in terms of their more known DPS counterpart: 
\begin{equation}
\label{eq:TPS_DPS_factor}
\sigmaefftps = k\times\sigmaeffdps, \; {\rm with}\;\; k = 0.82\pm 0.11\,.
\end{equation}
The relation is valid for different (typical) parton transverse profiles of proton.
In the numerical calculations below we take $\sigmaeffdps = 21$ mb which is rather a conservative choice but it corresponds to the average value extracted by the LHCb experiment only from the double charm data \cite{Aaij:2012dz}. This input gives us the value of $\sigmaefftps \simeq 17$ mb. 

In this paper, each of the single-parton scattering cross sections $\sigma_{pp \to c \bar c}^{\rm SPS}$ in Eq.~(\ref{doubleAB}) is calculated in the $k_{T}$-factorization approach \cite{kTfactorization} where higher-order QCD corrections are effectively included. In this framework
exact kinematics is kept from the very beginning and additional hard dynamics coming form transverse momenta of incident partons is taken into account.
It was shown in Ref.~\cite{Maciula:2013wg} that within this approach one can get a good description of the LHC inclusive charm data, similar to the case of next-to-leading order (NLO) collinear calculations. Likewise, the successful theoretical analyses of double charm production from Refs.~\cite{Maciula:2013kd,vanHameren:2014ava,Maciula:2016wci} were also based on the $k_{T}$-factorization.

According to this approach the differential SPS cross section for inclusive single $c\bar c$ pair production can be written as:             
\begin{eqnarray}
\frac{d \sigma_{pp \to c \bar c}^{\rm SPS}}{d y_1 d y_2 d^2 p_{1,t} d^2 p_{2,t}}
&& = \frac{1}{16 \pi^2 {(x_1 x_2 S)}^2} \int \frac{d^2 k_{1t}}{\pi} \frac{d^2 k_{2t}}{\pi}
\overline{|{\cal M}_{g* g* \rightarrow c \bar{c}}|^2} \nonumber \\
&& \times \;\; \delta^2 \left( \vec{k}_{1t} + \vec{k}_{2t} - \vec{p}_{1t} - \vec{p}_{2t}
\right)
{\cal F}_{g}(x_1,k_{1t}^2,\mu^2) {\cal F}_{g}(x_2,k_{2t}^2,\mu^2),
\label{ccbar_kt_factorization}
\end{eqnarray}
where the extra, compared to collinear factorization, integrals over transverse momenta $k_{it}$ of initial state particles appear. Here, ${\cal M}_{g^* g^* \rightarrow c \bar{c}}$ is the well-known gauge-invariant off-shell matrix element for $g^* g^* \to c \bar c$ partonic subprocess and ${\cal F}_{g}(x_i,k_{it}^2,\mu^2)$ are the so-called unintegrated (transverse momentum dependent) gluon PDFs (uPDFs).

The pocket-formula for TPS (Eq.~\ref{doubleAB}) can then be written in the differential form:   
\begin{equation} 
\label{pocket-kT}
\frac{d\sigma_{pp \to c \bar c c \bar c c \bar c }^{\rm TPS}}{d\xi_{12} \; d\xi_{34} \; d\xi_{56}} = \left(\frac{1}{3!}\right)\, \frac{1} {\sigmaefftps^2} \; \frac{d\sigma_{pp \to c \bar c}^{\rm SPS}}{d\xi_{12}} \cdot \frac{d\sigma_{pp \to c \bar c}^{\rm SPS}}{d\xi_{34}} \cdot \frac{d\sigma_{pp \to c \bar c}^{\rm SPS}}{d\xi_{56}},
\end{equation} 
where for simplicity $d\xi_{ij}$ stand for $d y_i d y_j d^2 p_{i,t} d^2 p_{j,t}$. 

In the present paper we use the Kimber-Martin-Ryskin (KMR) uPDFs \cite{KMR}, generated
from the LO set of a up-to-date Martin-Motylinski-Harland-Lang-Thorne (MMHT2014)
collinear gluon PDFs \cite{Harland-Lang:2014zoa} fitted also to the LHC data. In the perturbative part of the calculations we use
a running $\alpha_S^{LO}(\mu)$ provided with the MMHT2014 PDFs and the charm quark mass $m_{c} = 1.5$ GeV. We set both the renormalization and factorization scales equal to the averaged transverse mass $\mu^{2} = \frac{m^{2}_{c,t}+m^{2}_{\bar c,t}}{2}$, where $m_{c,t} = \sqrt{p^{2}_{c,t} + m^{2}_{c}}$.

The parton-level cross sections for triple charm quark (or charm antiquark) production are further corrected for the $c \to D$ (or $\bar c \to \bar{D}$) hadronization effects. This is done with the help of the fragmentation function (FF) technique. The quark cross section is transformed to 
$D$-meson hadron-level via the following procedure:   
\begin{equation}
\frac{d\sigma_{pp \to D D D }^{\rm TPS}}{d\xi^{D}}
 \approx
\int \frac{D_{c \to D}(z_{1})}{z_{1}}\cdot \frac{D_{c \to D}(z_{2})}{z_{2}} \cdot \frac{D_{c \to D}(z_{3})}{z_{3}}\cdot
\frac{d\sigma_{pp \to c c c }^{\rm TPS}}{d\xi^{c}} d z_{1} d z_{2} d z_{3} \; ,
\end{equation}
where $d\xi^{a}$ stand for $d y^{a}_1 d y^{a}_2 d y^{a}_3 d^2 p^{a}_{1,t} d^2 p^{a}_{2,t} d^2 p^{a}_{3,t}$ taking $a = c$ quark or $D$ meson and
$p^{c}_{i,t} = \frac{p_{i,t}^{D}}{z_{i}}$ with meson momentum fractions  $z_{i}\in (0,1)$.
The usual approximation here is that the quark rapidities $y^{c}_{1}, y^{c}_{2}, y^{c}_{3}$ are unchanged in the fragmentation process which is known
to be especially legitimate in the case of heavy flavors and meson transverse momenta larger than its mass (see \textit{e.g.} Ref.~\cite{Maciula:2015kea}). In the numerical calculations here we use the commonly used in the literature scale-independent Peterson FF \cite{Peterson:1982ak} with the parameter $\varepsilon_{c} = 0.05$, which is the averaged value extracted from different $e^{+}e^{-}$ experiments. In the last step the obtained cross sections for triple-meson production in the way sketched above are normalized with the corresponding
fragmentation fraction $\mathrm{BR}(c \to D^{0}) = 0.565$ \cite{Lohrmann:2011np}.

\section{Numerical results}

We start with our predictions for multiple $c \bar c$ production.
In order to compare our results to the results of Ref.~\cite{dEnterria:2016ids}
in Table~\ref{tab:total-cross-sections} we show cross sections for the full phase
space for two different collision energies. We get considerably larger
cross section for triple $c \bar c$ production compared to the numbers
read off from Fig.~1 of Ref.~\cite{dEnterria:2016ids}. It is not obvious how to understand
the difference, as rather different approaches were used in both cases. There are obvious uncertainties
related to the choice of factorization/renormalization scales in both approaches.
In our approach the region of very small transverse momenta of $c$ or $\bar{c}$ quarks is the least certain, 
which is related to uncertain region of very small gluon transverse momenta. Small uncertainties for single-parton scattering
are considerably magnified for triple-parton scattering. Large differences of cross section in the full phase space
do not necessarily involve such differences for fiducial volume. We think that agreement of theoretical inclusive cross sections
for $D$ meson production in a fiducial volume is a necessary (and sufficient) condition for the best estimating
of DPS and TPS effects.   

\begin{table}[tb]%
\caption{Total charm SPS, DPS and TPS cross sections (in mb) in $pp$-collisions at the LHC, calculated in the $k_{T}$-factorization approach and using the factorized Ansatz.}
\label{tab:total-cross-sections}
\centering %
\newcolumntype{Z}{>{\centering\arraybackslash}X}
\begin{tabularx}{1.0\linewidth}{Z Z Z}
\toprule[0.1em] %

Final state  & $\sqrt{s} = 7$ TeV  & $\sqrt{s} = 13$ TeV       \\ [-0.2ex]

\bottomrule[0.1em]

SPS: $\sigma(c\bar{c} + X)$  & 9.84  & 17.30 \\  [-0.2ex]
DPS: $\sigma(c\bar{c}c\bar{c} + X)$  & 2.31  & 7.13 \\  [-0.2ex]
TPS: $\sigma(c\bar{c}c\bar{c}c\bar{c} + X)$  & 0.55  & 2.99 \\  [-0.2ex]

\bottomrule[0.1em]

\end{tabularx}
\end{table}

\begin{figure}[!h]
\begin{minipage}{0.47\textwidth}
 \centerline{\includegraphics[width=1.0\textwidth]{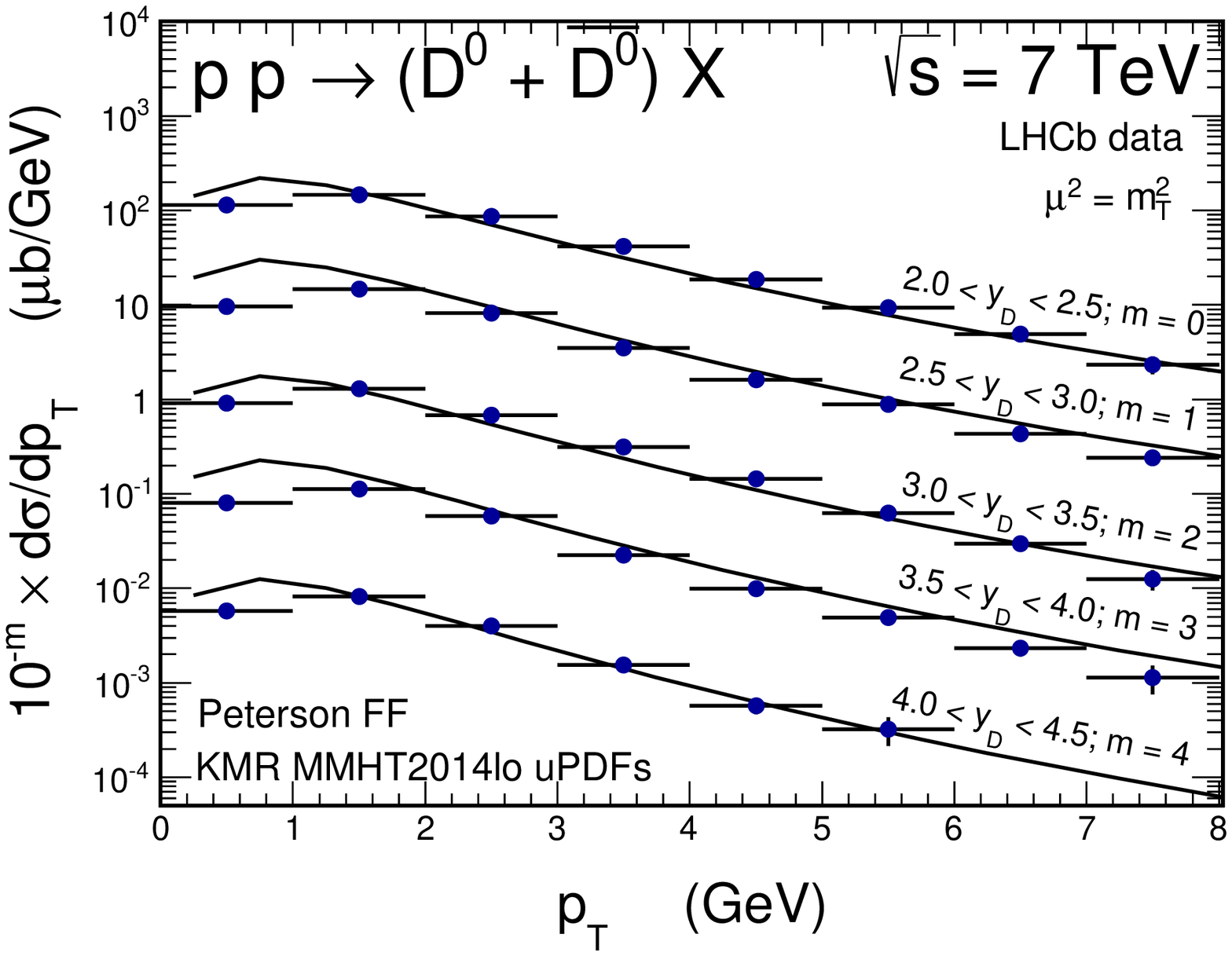}}
\end{minipage}
\hspace{0.5cm}
\begin{minipage}{0.47\textwidth}
 \centerline{\includegraphics[width=1.0\textwidth]{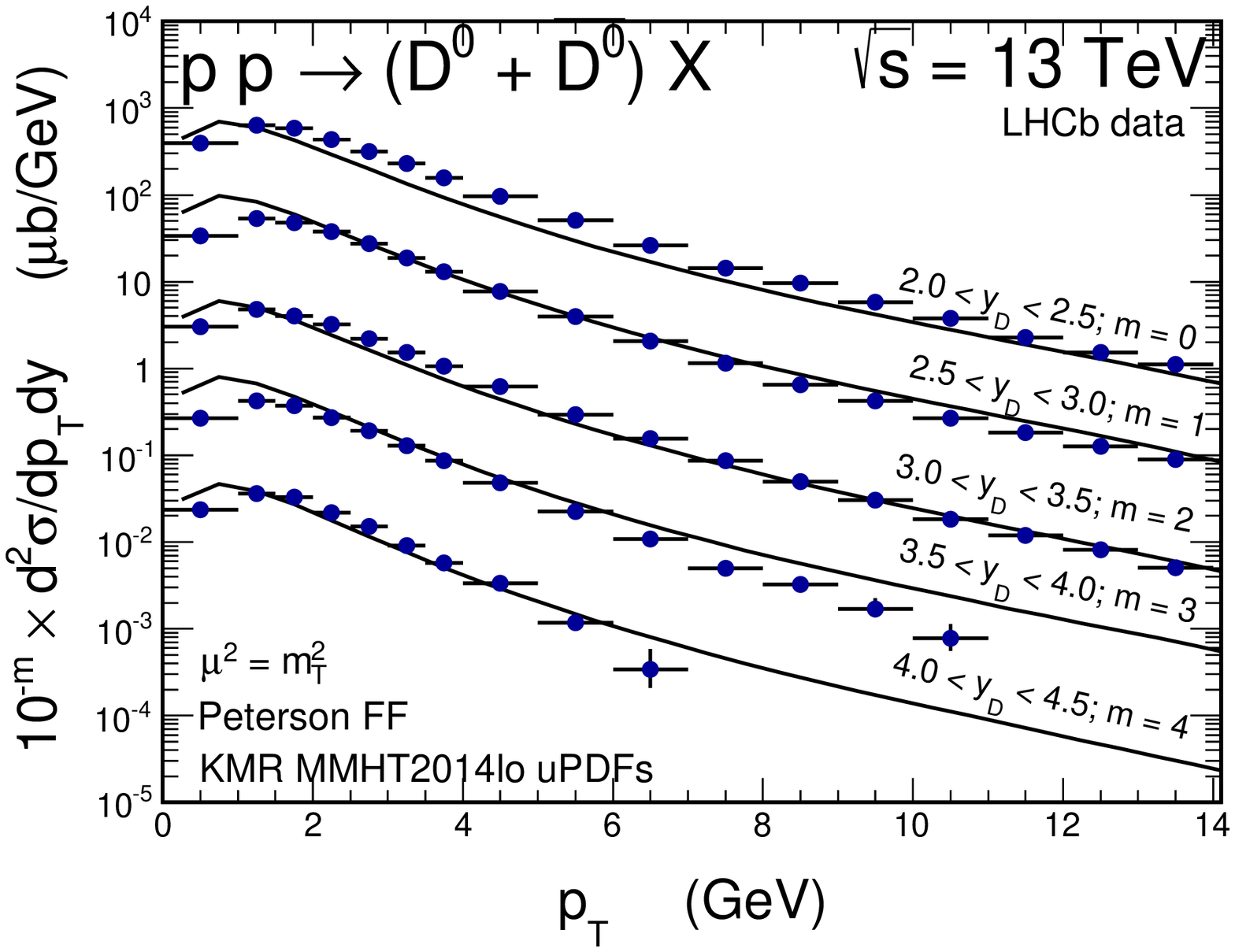}}
\end{minipage}
   \caption{
\small Transverse momentum distributions of $D^{0}$ meson in the case of inclusive single meson production measured by the LHCb experiment at
 $\sqrt{s}=7$ TeV \cite{Aaij:2013mga} (left panel) and $\sqrt{s}=13$ TeV \cite{Aaij:2015bpa} (right panel). The solid lines correspond to our theoretical predictions based on the $k_{T}$-factorization approach with the KMR uPDFs. The results and experimental data points are shown for different rapidity bins defined in the figure. 
 }
 \label{fig:inclusive-D0}
\end{figure}

In Fig.~\ref{fig:inclusive-D0} we present our results for
inclusive single $D^0$ meson production for the LHCb experiment
at $\sqrt{s}$ = 7 and 13 TeV. In both cases we get good description of 
the measured transverse momentum distributions for different
intervals of rapidity. The good agreement is a good starting point
for calculation of double and triple $D$ meson production.

In Table \ref{tab:total-cross-sections-D0} we show our predicted cross
sections for two and three mesons within the fiducial volume
of the LHCb detector. The $DD$ pairs were already measured by
the LHCb collaboration. The predicted value at $\sqrt{s}$ = 7 TeV is
consistent with the measured one (see Table 12 in Ref.~\cite{Aaij:2012dz}).
Whether the three mesons can be measured at the LHCb will be discussed 
in the following.

\begin{table}[tb]%
\caption{The integrated cross sections for double and triple $D^{0}$ meson production (in nb) within the LHCb acceptance: $2 < y_{D^{0}} < 4$ and $3 < p_{T}^{D^{0}} < 12$ GeV, calculated in the $k_{T}$-factorization approach. The numbers include also the charge conjugate states.}
\label{tab:total-cross-sections-D0}
\centering %
\newcolumntype{Z}{>{\centering\arraybackslash}X}
\begin{tabularx}{1.0\linewidth}{Z Z Z}
\toprule[0.1em] %

Final state  & $\sqrt{s} = 7$ TeV  & $\sqrt{s} = 13$ TeV       \\ [-0.2ex]

\bottomrule[0.1em]

DPS: $\sigma(D^{0}D^{0} + X)$  & 784.74  & 2992.91 \\  [-0.2ex]
TPS: $\sigma(D^{0}D^{0}D^{0} + X)$  & 2.38  & 17.71 \\  [-0.2ex]

\bottomrule[0.1em]

\end{tabularx}
\end{table}

\begin{figure}[!h]
\begin{minipage}{0.47\textwidth}
 \centerline{\includegraphics[width=1.0\textwidth]{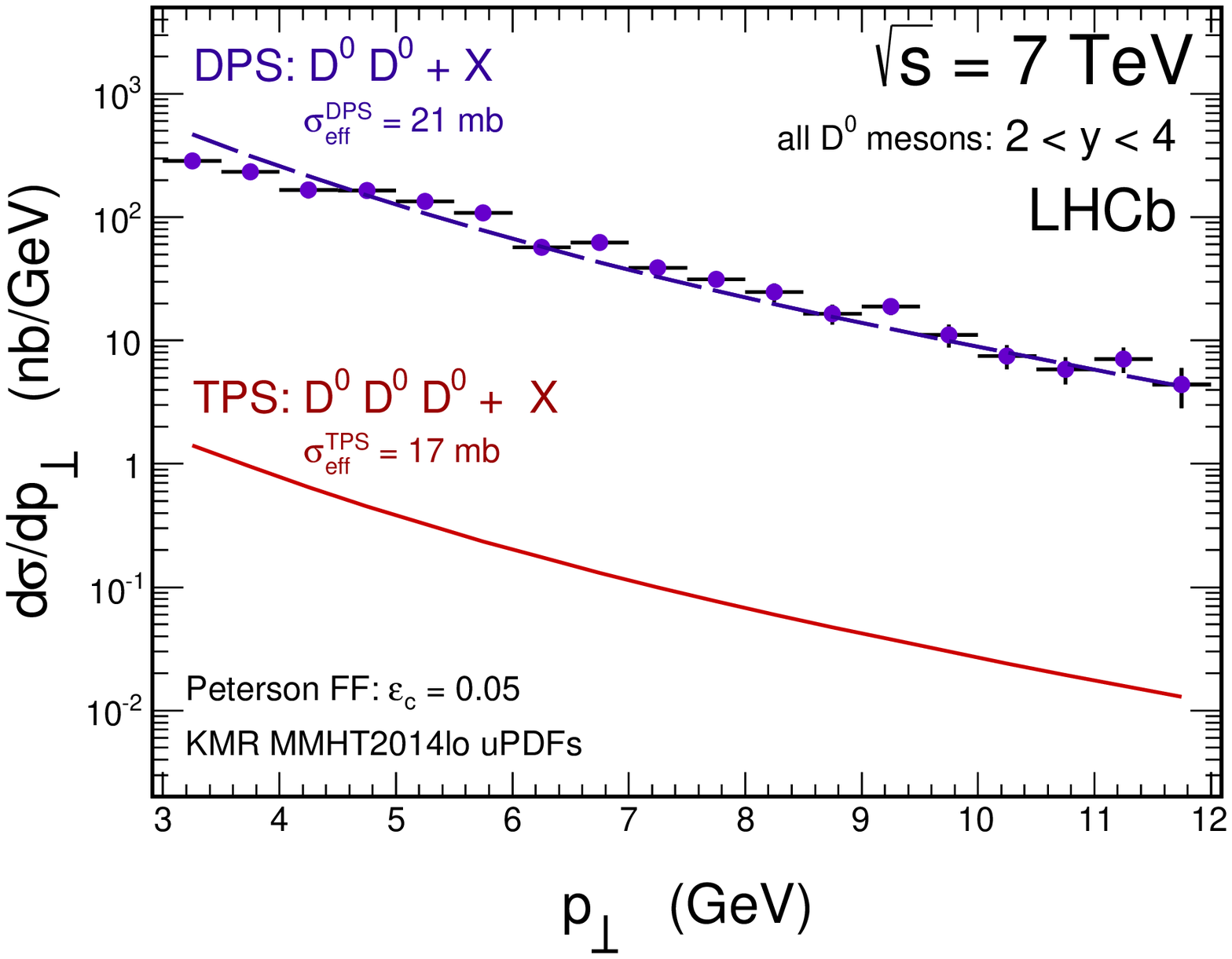}}
\end{minipage}
\hspace{0.5cm}
\begin{minipage}{0.47\textwidth}
 \centerline{\includegraphics[width=1.0\textwidth]{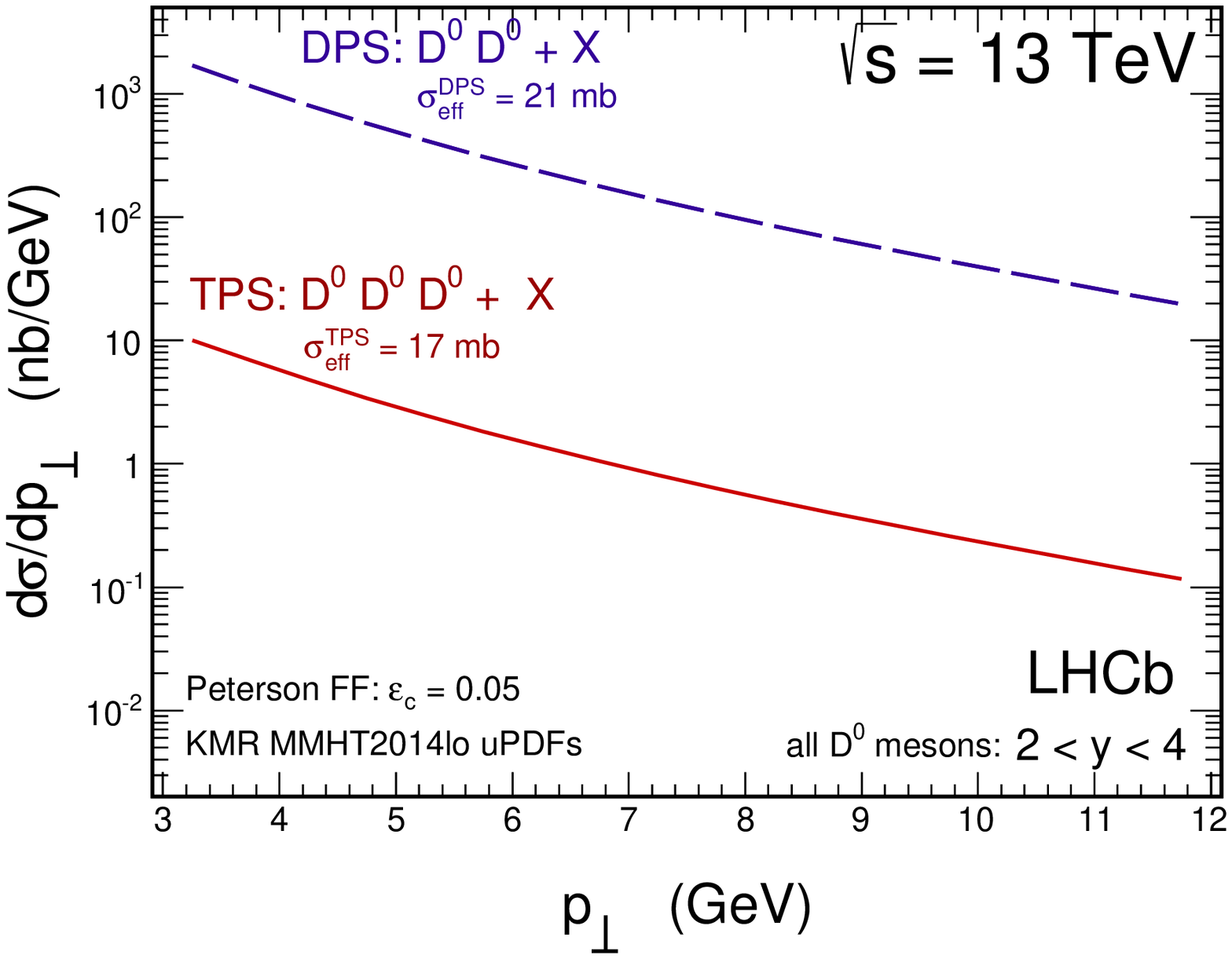}}
\end{minipage}
   \caption{
\small Transverse momentum distributions of one of the measured $D^{0}$ mesons for double-$D^{0}$ (upper long-dashed lines) and triple-$D^{0}$ (lower solid lines) production for $\sqrt{s} = 7$ (left) and $13$ TeV (right). Details are specified in the figure. The data points for double-$D^{0}$ production in the left panel are taken from Ref.~\cite{Aaij:2012dz}.
 }
 \label{fig:double-triple-D0}
\end{figure}

Now we wish to discuss some differential distributions for
double and triple $D^0$ meson production.
In Fig.~\ref{fig:double-triple-D0} we show transverse momentum distribution of one 
of the two or one of the three $D^0$ mesons (all measured by the LHCb detector).
In the used multiple parton scattering formalism the distributions
for single, double and triple production have the same shape/slope
and differ only by normalization. The distributions
for triple $D^0$ production are about two orders of magnitude smaller
than for double $D^0$ production, consistent with Table \ref{tab:total-cross-sections-D0}.
In the left panel, for $\sqrt{s}$ = 7 TeV, we also show for reference
the LHCb experimental data \cite{Aaij:2012dz}. As for single scattering case
(see Fig.\ref{fig:inclusive-D0}) we get a good agreement also here, so we hope 
our predictions for triple $D^0$ production (lowest curves) are (should be) reliable.

\begin{figure}[!h]
\begin{minipage}{0.47\textwidth}
 \centerline{\includegraphics[width=1.0\textwidth]{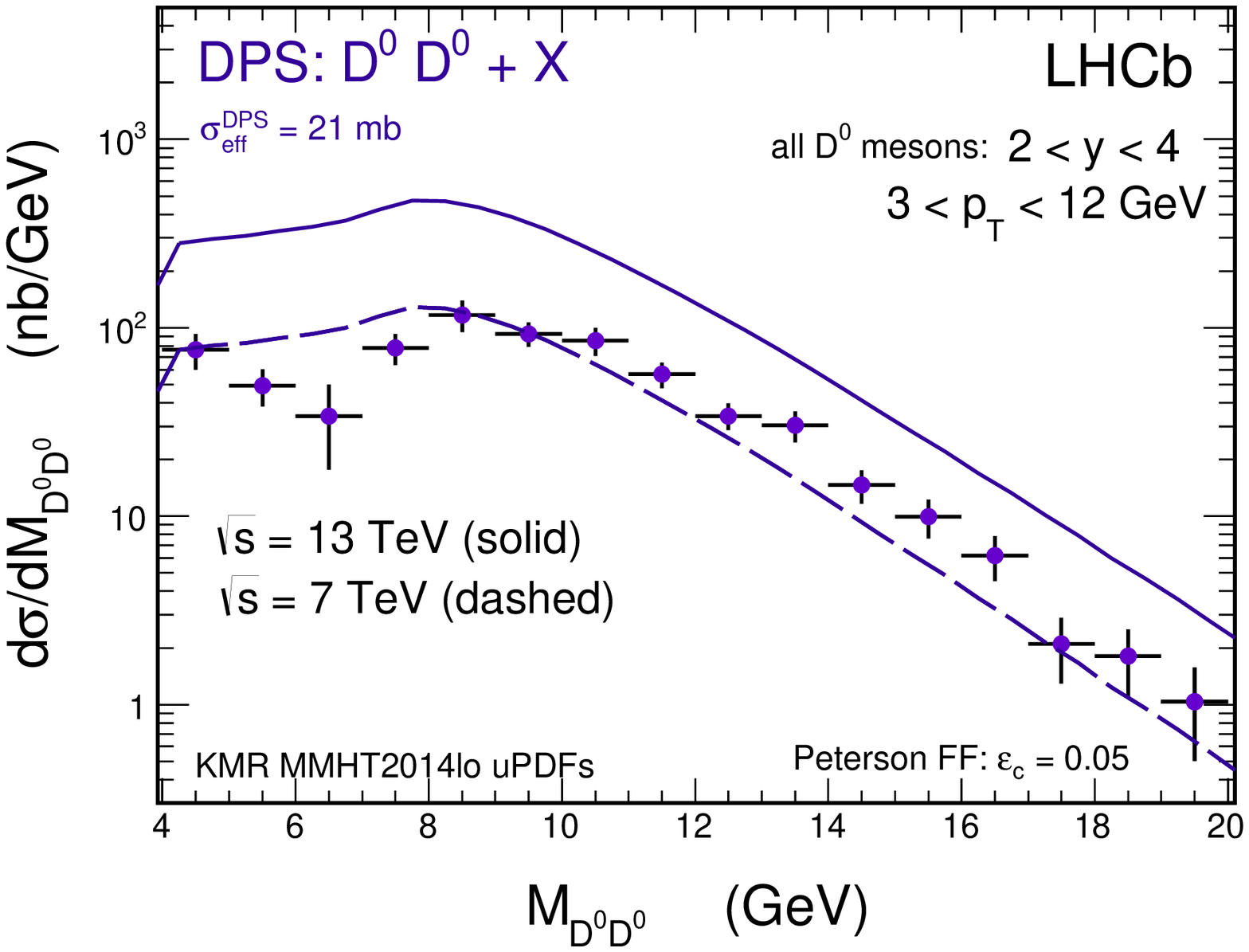}}
\end{minipage}
\hspace{0.5cm}
\begin{minipage}{0.47\textwidth}
 \centerline{\includegraphics[width=1.0\textwidth]{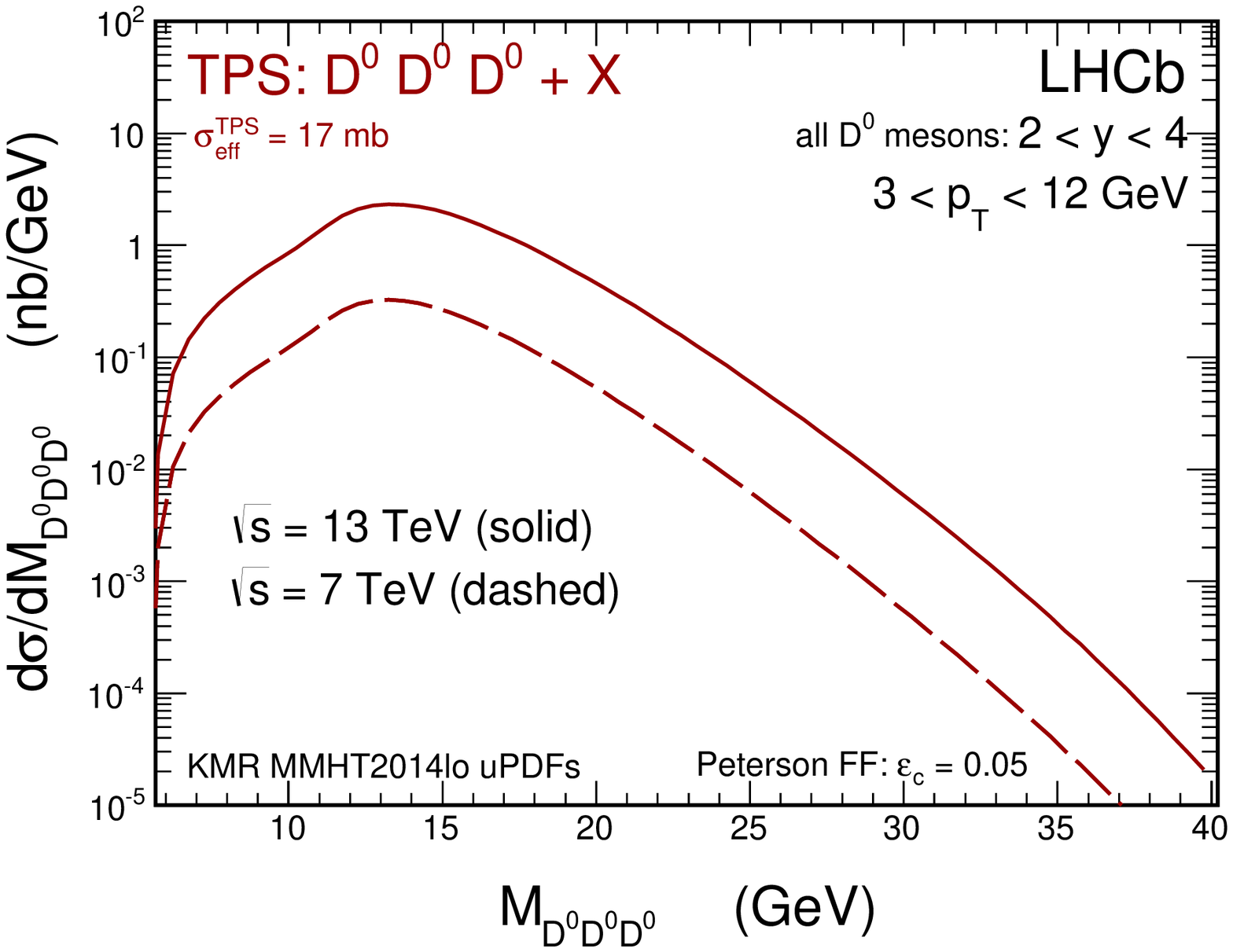}}
\end{minipage}
   \caption{
\small Invariant mass distributions (corresponding to the LHCb acceptance) of the di-meson $D^{0}D^{0}$ system for the DPS mechanism (left) and tri-meson $D^{0}D^{0}D^{0}$ system for the TPS mechanism (right), for $\sqrt{s} = 7$ TeV (lower long-dashed lines) and $13$ TeV (upper solid lines). Details are specified in the figure. The data points for double-$D^{0}$ production in the left panel are taken from Ref.~\cite{Aaij:2012dz}.
 }
 \label{fig:double-triple-D0-Minv}
\end{figure}

In Fig.\ref{fig:double-triple-D0-Minv} we show our predictions
for invariant mass distributions for two (left) and three (right)
$D^0$ mesons. The lower curves are for $\sqrt{s}$ = 7 TeV and the upper
curves for $\sqrt{s}$ = 13 TeV. Again we show the LHCb experimental
data for $\sqrt{s}$ = 7 TeV (left panel).
The distributions shown in the right panel are waiting for experimental
verification.

Finally in Table \ref{tab:events} we show the number of counts
for different realistic values of the integrated luminosity for the LHCb experiment. The predicted numbers of events for double- and triple-$D^{0}$
production correspond to the cross sections from Table~\ref{tab:total-cross-sections-D0}. Here we have included
in addition the relevant decay branching fraction $\mathrm{BR}(D^{0} \to K^{-} \pi^{+}) = 0.0393$ \cite{Olive:2016xmw}.

\begin{table}[tb]%
\caption{Number of events for different values of the feasible integrated luminosity in the LHCb experiment for the calculated cross sections from Table~\ref{tab:total-cross-sections-D0}. The branching fractions for $D^{0} \to K^{-}\pi^{+} (\bar{D^{0}} \to K^{+}\pi^{-})$ are included here.}
\label{tab:events}
\centering %
\newcolumntype{Z}{>{\centering\arraybackslash}X}
\begin{tabularx}{1.0\linewidth}{Z Z Z Z}
\toprule[0.1em] %

$\sqrt{s}$ & Integrated Luminosity & DPS ($D^{0}D^{0}$) & TPS ($D^{0}D^{0}D^{0}$) \\

\bottomrule[0.1em]

 \multirow{2}{1.2cm}{$7$ TeV} &  355 pb$^{-1}$   &  $0.43 \times 10^{6}$   & 51    \\
                             &  1106 pb$^{-1}$  &  $1.34 \times 10^{6}$    & 159     \\
\hline
 \multirow{2}{1.2cm}{$13$ TeV} &  1665 pb$^{-1}$   &  $7.70 \times 10^{6}$ & 1789    \\
                             &  5000 pb$^{-1}$  &  $23.11 \times 10^{6}$   & 5374     \\                             


\bottomrule[0.1em]

\end{tabularx}
\end{table}

\section{Conclusions}

In this letter we have made first estimation of the cross sections for
triple $D^0$ production within the LHCb fiducial volume
in order to verify triple parton scattering effects for triple 
$c \bar c$ production. 

We have obtained rather large cross sections for triple $c \bar c$
production, larger than predicted very recently in
Ref.~\cite{dEnterria:2016ids} but still consistent within uncertainties of the two approaches. We have checked, however, our approach against 
inclusive single $D^0$ and double $D^0 D^0$ production as measured 
by the LHCb collaboration. In both cases we have obtained a 
fairly good agreement which gives us confidence for triple 
$D^0$ production.

We have presented both integrated cross section as well as
differential distributions for double and triple $D^0$ production
for $\sqrt{s}$ = 7 and 13 TeV.
We have presented also predicted number of counts for different realistic values of
integrated luminosity for the LHCb experiment.
In the case of triple $D^0$ production we have predicted about 100 counts at $\sqrt{s}= 7$ TeV and a few thousands of counts at $\sqrt{s}= 13$ TeV
for realistic integrated luminosities. 
We hope the LHCb collaboration will be able to verify our
predictions soon.

\vspace{1cm}

{\bf Acknowledgments}

This study was partially
supported by the Polish National Science Center grant
DEC-2014/15/B/ST2/02528 and by the Center for Innovation and
Transfer of Natural Sciences and Engineering Knowledge in
Rzesz{\'o}w.


\end{document}